\title{PCEB paper}
\documentclass{aa}
\usepackage{graphicx}
\usepackage{txfonts}
\usepackage{natbib,twoopt}
\usepackage{hyperref} 
\bibpunct{(}{)}{;}{a}{}{,} 
\begin{document}

\title{Applegate mechanism in post-common-envelope binaries: Investigating the role of rotation}
\titlerunning{The Applegate mechanism in PCEBs: Investigating the role of rotation}
   \author{F.H. Navarrete\inst{1}
          \and
          D.R.G. Schleicher\inst{1}
          \and
          J. Zamponi Fuentealba\inst{1}
          \and
           M. V\"olschow\inst{2}
          }

   \institute{Departamento de Astronom\'ia, Facultad de Ciencias F\'isicas y Matem\'aticas, Universidad de Concepci\'on, Av. Esteban Iturra s/n, Barrio Universitario, Casilla 160-C, Chile \email{felnavarrete@udec.cl} \and Hamburger Sternwarte, Universit\"at Hamburg, Gojenbergsweg 112, 21029 Hamburg, Germany              
             }

   \date{Received -; accepted -}
 
  \abstract
   {Eclipsing time variations are observed in many close binary systems. In particular, for several post-common-envelope binaries (PCEBs) that consist of a white dwarf and a main sequence star, the Observed minus Calculated (O-C) diagram suggests that real or apparent orbital period variations are driven by Jupiter-mass planets or as a result of magnetic activity, the so-called Applegate mechanism. The latter explains orbital period variations as a result of changes in the stellar quadrupole moment due to magnetic activity.}
   {In this work we explore the feasibility of driving eclipsing time variations via the Applegate mechanism for a sample of PCEB systems, including a range of different rotation rates.}
   {We used the MESA code to evolve 12 stars with different masses and rotation rates. We applied simple dynamo models to their radial profiles to investigate the scale at which the predicted activity cycle matches the observed modulation period, and quantifiy the uncertainty. We further calculated the required energies to drive the Applegate mechanism.}
   {We show that the Applegate mechanism is energetically feasible in 5 PCEB systems. In RX~J2130.6+4710, it may be feasible as well considering the uncertainties. We note that these are the systems with the highest rotation rate compared to the critical rotation rate of the main-sequence star.}
   {The results suggest that the ratio of physical to critical rotation rate in the main sequence star is an important indicator for the feasibility of Applegate's mechanism, but exploring larger samples will be necessary to probe this hypothesis.}


\maketitle

\section{Introduction}
Post-common-envelope binaries (PCEBs) are systems that consist of a low-mass main-sequence secondary and a white dwarf (WD). Their evolution, first proposed by \citet{paczynski76}, starts when the more massive component evolves up to a stage where its surface goes beyond the outer Lagrangian point, thus engulfing its companion. At this point the less massive star experiences friction as it orbits and thus loses orbital energy and angular momentum to the common envelope, spiraling inwards until enough energy is transfered to the envelope for it to be expelled. This leaves a close binary consisting of an M dwarf (dM) or a subdwarf and a white dwarf, with a binary period of normally less than three hours \citep[see e.g.,][]{parsons13}.

\indent Besides the common envelope evolution itself, these systems reveal an intriguing characteristic when the observed minus calculated (O-C) eclipse times diagram is constructed. Periodic variations on timescales of a few years to a few tens of years suggest the existence of giant planets orbiting the binary \citep[see e.g.,][]{beuermann10, nasiroglu17}. There are two possibilities to explain the presence of these hypothetical planets. They might have formed together with the binary and then survived the common envelope phase (first-generation planet formation), or they could have formed from the common envelope material (second-generation planet formation). A hybrid scenario may be also possible, with accretion of the ejected gas onto already existing planets. Several studies have been carried out on this matter and some authors favor one scenario over the other \citep[see e.g.,][]{voelschow14, schleicher14, bear14, schleicher15}.
In any case, these scenarios must take into account that $\sim$90\% of the PCEBs, as found by \citet{zorotovic13}, have observed apparent period variations, implying that the hypothetical planets are very good at surviving the PCEB formation process and/or the common envelope material is very efficient in forming planets.

\indent Another possible explanation that does not invoke the presence of planets is the so-called Applegate mechanism \citep{applegate92}. In this scenario, the magnetic activity of the M dwarf redistributes the angular momentum within the star, thus changing its gravitational quadrupole moment (i.e., the shape of the star). This in turn produces a variation on the binary separation which translates into variations of the O-C diagram. The original Applegate model has been improved by several authors. For instance, \citet{Lanza99} improved the model by adopting a consistent description for stellar virial equilibrium \citep{Chandrasekhar61}, while \citet{Brinkworth06} extended the model by introducing a finite shell formalism, considering the exchange of the angular momentum between the shell and the core. This model was further examined and systematically applied by \citet{voelschow16} to a sample of 16 close binary systems, involving predominantly PCEBs, showing that the Applegate mechanism is a viable process in the shortest and most massive binary systems. 

As they discuss, many of the PCEBs indeed show signs of magnetic activity. For instance in the case of  V471 Tau, magnetic activity was inferred via photometric variability, flaring events and H$\alpha$ emission along with a strong X-ray signal \citep{Kaminski07, Pandey08}. In case of DP~Leo, magnetic activity was shown through X-ray observations \citep{Schwope02}, while for QS~Vir, it has been inferred via detections of Ca~II emission and Doppler imaging \citep{Ribeiro10} and observed coronal emission \citep{Matranga12}. In the case of HR~1099, a 40 year X-ray light curve suggesting a long-term cycle was recently compiled by \citet{Perdelwitz17}, and similar studies have been pursued via optical data \citep[e.g.,][]{Donati03, Lanza06, Berdyugina07, Muneer10}. It is also worth mentioning that in case of V471~Tau, the possibility of a third body has been ruled out via direct imaging observations \citep{Hardy15}.

In this paper, our goal is to further assess whether magnetic activity is a feasible alternative to the presence of a third body, at least in some of the systems. In particular, we note that in the context of  isolated stars, characteristic relations have been inferred between the activity cycle and the rotation period \citep{Saar99, Boehm07}, which can be interpreted in terms of simple dynamo models, as put forward by \citet{Soon93} and \citet{Baliunas96}. In this study we have employed these models to estimate the expected duration of the activity cycle for a sample of PCEBs, and compared it to the observed modulation periods. In addition, we re-assess the feasibility of the Applegate mechanism using the model of \citet{voelschow16}, using improved stellar structure parameters for the individual systems.

The structure of this paper is as follows: In Section~\ref{simulations}, we describe the stellar structure calculations to derive the stellar profiles, our dynamo model as well as the system of equations required to assess the feasibility of the Applegate mechanism. Our results are presented in Section~\ref{results}, and a comparison with selected systems is given in Section~\ref{comparison}. The final discussion and conclusions are presented in Section~\ref{conclusions}.

\section{Simulations and calculations}\label{simulations}

In the following, we describe the MESA star calculations as well as the analytical framework employed in this paper.

\subsection{MESA star calculations}
We used the Modules for Experiments in Stellar Astrophysics \citep[MESA;][]{mesa11} code\footnote{Webpage MESA: http://mesa.sourceforge.net/} to simulate the secondary star for every system under consideration. In particular we used the module MESA star, which is a one-dimensional stellar evolution code to evolve a single star. Within this code, we focused on three main input parameters of interest for our simulation, which are:

\begin{enumerate} 

\item Stellar mass: Conserved during the simulation and equal to the mass of the convective star in each system.
\item Rotation: Set as the fraction $\Omega_{\textnormal{ZAMS}} / \Omega_{\textnormal{crit}}$, where $\Omega_{\textnormal{crit}}~=~\sqrt{GM/R^3}$ is the critical angular velocity, above this limit the star breaks up. $\Omega_{\textnormal{ZAMS}}$ is the angular velocity of the star at the zero age main sequence (ZAMS). We chose to evolve the stars at a fixed $\Omega_{\textnormal{ZAMS}} / \Omega_{\textnormal{crit}}$ during all the simulation. Hereafter we drop the \textit{ZAMS} subscript
\item Maximum age. We stopped the simulation when the star reaches the derived system age or radius.
\end{enumerate}

\noindent We used this last profile, at its maximum age, for our calculations. For stars with masses lower than $0.14\,M_\odot$ the very low mass test-suite available in MESA was used as a guide to properly evolve these systems. For every run we chose the 'ML1' mixing length implementation with the free parameter $\alpha_{\textnormal{MLT}} = 1.5$ and the basic reaction network. The calculations were pursued assuming solar metallicity. We note that \citet{voelschow16} shows that the resulting stellar profiles are rather insensitive to the precise value of the age or metallicity in the case of low-mass stars.

\subsection{Analytical dynamo models}

The MESA models are used as input for the stellar structure to explore the expectations based on simple dynamo models. We employed here a similar approach to that of \citet{schleichermennickent17} and \citet{Perdelwitz17}, using the relation between the period of the cycle and the period of rotation \citep{Soon93, Baliunas96}
\begin{equation}\label{eq:period}
P_\textnormal{cycle} = D^\gamma\,P_\textnormal{rot},
\end{equation}
with $P_\textnormal{cycle}$ the activity cycle of the star, $D$ the dynamo number, and $P_\textnormal{rot}$ the rotation period of the star. As the PCEBs are close binaries, we were able to assume the secondary star is tidally locked \citep{Zahn89a, Zahn89b}, making $P_\textnormal{rot}$ equal to the binary period. $\gamma$ is a parameter that depends on the level of activity of the star \citep{dube13}. $D$ is further related to the Rossby number $Ro$ via $D=$ Ro$^{-2}$, where Ro $= P_\textnormal{rot}/\tau_c$ is the ratio between the stellar rotation period $P_\textnormal{rot}$, and the convective turnover time, $\tau_c$. Following \citet{soker00}, we calculated Ro as
\begin{equation}\label{eq:ro}
\textnormal{Ro} = 9\left(\frac{v_c}{10\,\textnormal{km/s}}\right)\left(\frac{H_p}{40\,R_\odot}\right)^{-1}\left(\frac{\omega}{0.1\,\omega_\textnormal{Kep}}\right)^{-1}\left(\frac{P_\textnormal{Kep}}{\textnormal{yr}}\right).
\end{equation}
Here, $v_c$ and $H_p$ are the convective velocity and pressure scale height, respectively. Both are stellar properties given in the MESA output files. $\omega$ is the angular velocity, $\omega_\textnormal{Kep}$ and $P_\textnormal{Kep}$ are the Keplerian angular velocity and orbital period, respectively. These last three values are obtained from the binary parameters.

As described by \citet{Perdelwitz17}, we first simulated a Sun-like star to obtain $\gamma$ in (\ref{eq:period}) taking into consideration that the tachocline at $r = 0.7\,R_\odot$ is where the dynamo is expected to give rise to a strong toroidal magnetic field with a cycle of $22$~years. From this calibration, we obtained a reference value of $\gamma = 0.86$. However, the Sun has a rather slow rotation rate, with a period of about $24.5$ days at the equator, and a relatively weak magnetic field of about $\sim1$~G. Even single M-dwarfs, on the other hand, have rotation periods of less than ten days \citep{somers17}, and in PCEBs rotation periods of less than three hours, with field strengths on the order of a few kG \citep{johns-krull96}. As a result, the actual value of $\gamma$ may be different, and we thus also explored $\gamma=0.7$ and $\gamma=1.0$ in Table~\ref{table:alpha}. For a given value of $\gamma$, it is then possible to calculate the fraction of the stellar radius $R_d/R$ where the predicted activity period is equal to the observed one.

\subsection{The Applegate mechanism}

Given an idea of the radius where the dynamo operates inside the star, it is now also important to find out whether it is energetically feasible to produce the observed eclipsing time variations via magnetic activity. We employed the Applegate mechanism in the formulation presented by \citet{voelschow16}, which we recently implemented in the now publicly available  \texttt{Applegate calculator}\footnote{Applegate calculator:\newline http://theory-starformation-group.cl/applegate/}. The latter in particular allows us to calculate the required energy $\Delta E$ as a fraction of the available energy in the magnetically active star, $E_\textnormal{sec}$, to drive the corresponding change of the quadrupole moment. The equation is given as
\begin{equation}\label{eq:energy}
\frac{\Delta E}{E_{sec}} = k_1\,\frac{M_\textnormal{sec}R_\textnormal{sec}^2}{P_\textnormal{bin}^2P_\textnormal{mod}L_{sec}}\left(1\pm\sqrt{1-k_2G\frac{a_\textnormal{bin}^2M_\textnormal{sec}P_\textnormal{bin}^2}{R_\textnormal{sec}^5}\frac{\Delta P}{P_\textnormal{bin}}}\right)^2,
\end{equation}
where $P_\textnormal{bin}$, $a_\textnormal{bin}$, and $\Delta P/P_\textnormal{bin}$ are the orbital period, semi-major axis, and observed relative change of the orbital period during one cycle of the binary, respectively; $P_\textnormal{mod}$ is the observed modulation period of the binary; $M_\textnormal{sec}$, $R_\textnormal{sec}$, $L_\textnormal{sec}$, $T_\textnormal{sec}$ are the mass, radius, luminosity, and temperature of the magnetically active star, respectively. The $k_1$ and $k_2$ parameters are defined as
\begin{align}
 k_1 &= \frac{4\pi^2}{5}\frac{\lambda(\gamma+1)(\xi^3-\xi^{-2})}{1+\lambda(\xi^3-1)}\frac{(f-\gamma)^2}{(\gamma^2/\lambda+f)^2}, \\
 k_2 &= \frac{15}{36\pi^2}\frac{\xi^5}{\lambda}\frac{(\gamma/\lambda+f)}{(f-\gamma)^2}.
\end{align}
These parameters depend on the stellar structure. The first parameter regulates the overall magnitude of the required energy to drive quasi-periodic period variations, and the second affects the separation between low- and high-energy solution for the Applegate mechanism. Both $k_1$ and $k_2$ depend on two mean densities inside the star: $\rho_{\rm in}$ from the bottom to $R_d$, and $\rho_{\rm out}$ from $R_d$ to $R_\textnormal{star}$. The model assumes interchange of angular momentum between these two zones.The additional parameters $\lambda$ and $\xi$ are defined as $\lambda = \rho_\textnormal{out} / \rho_\textnormal{in}$ and $\xi = R_\textnormal{star} / R_d$.

An important parameter here is $R_{\rm core}$, the radius that separates the core and the shell between which angular momentum is exchanged. As a first approach, we assumed that it corresponds approximately to the scale where the dynamo model produces an activity cycle that corresponds to the observed period. We however also discuss other choices and explore their implications. We note that we explicitly evaluated here the parameters $k_1$ and $k_2$ based on the stellar profiles obtained from MESA. As also discussed by \citet{Lanza06}, the angular velocity variations in the radiative core are confined to a thin layer, so its contribution to the changes on the quadrupole moment are negligible. We therefore have chosen to assume that the interchange of angular momentum is confined to the convective zone, thus neglecting any contribution of the radiative cores found in the secondaries of QS~Vir and V471~Tau. The latter was taken into account when calculating the mean density in the core and thus in the resulting stellar structure parameters.

\begin{figure}
\resizebox{\hsize}{!}{\includegraphics{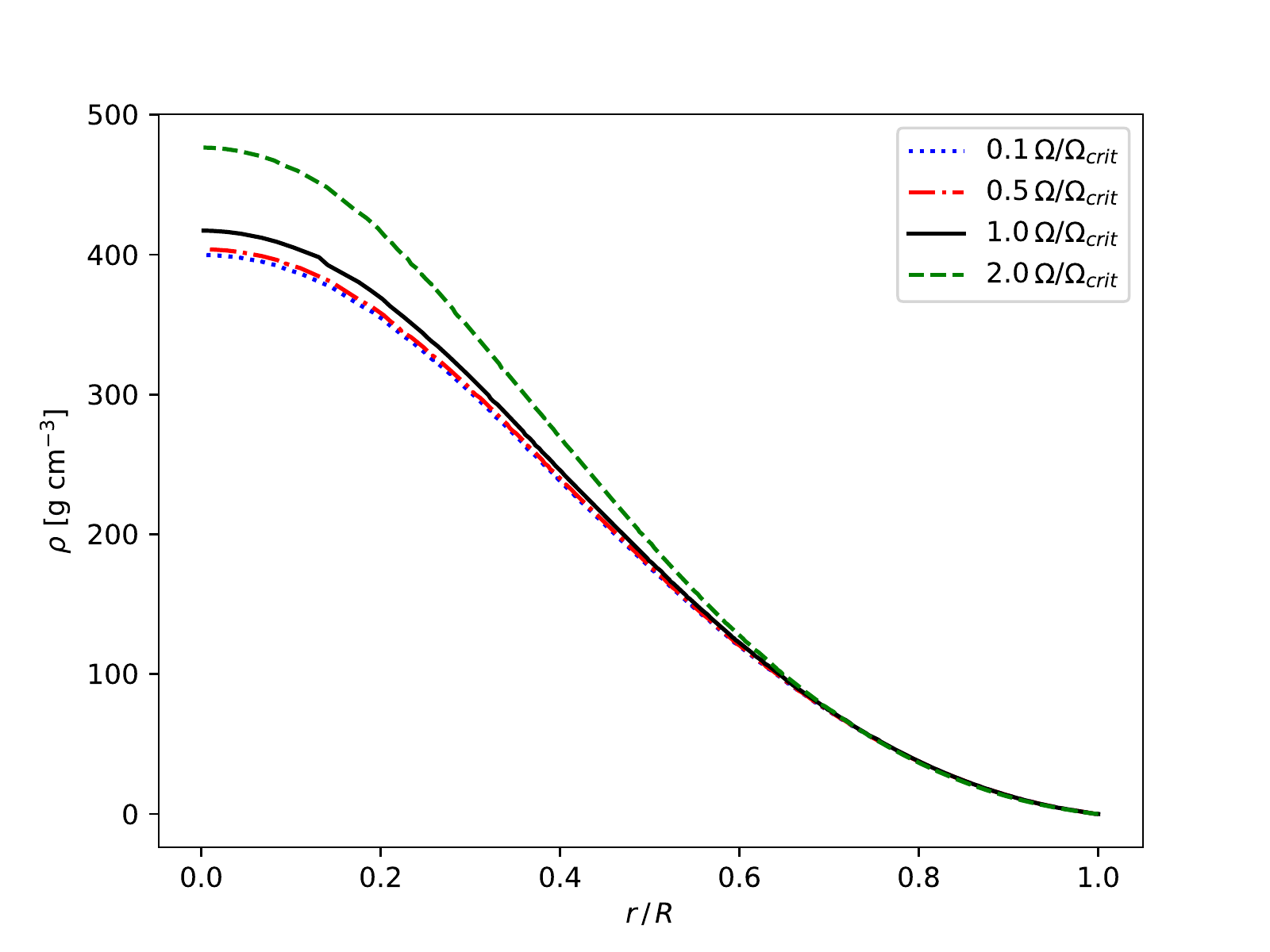}}
\resizebox{\hsize}{!}{\includegraphics{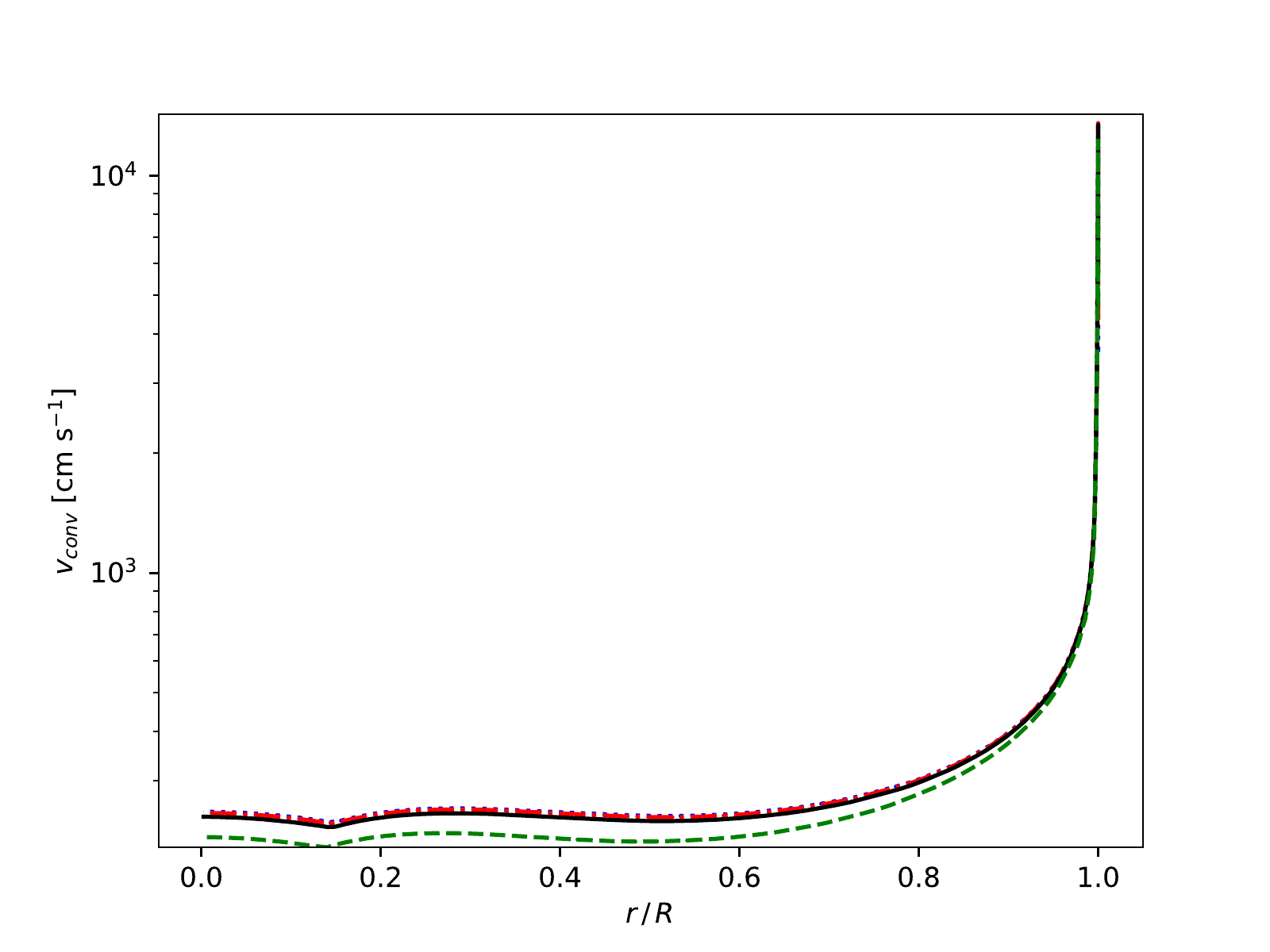}}
\caption{Density profile (top) and convection velocity profile (bottom) for DP~Leo simulated by MESA for different rotation velocities. The ratio $1.0\,\Omega/\Omega_\textnormal{crit}$ corresponds to the observed one and is $\sim$~$0.288$ for DP~Leo.}
\label{fig:dp-leo-conv-vel}
\end{figure}

\section{Results}\label{results}
In the following, we present the results from our stellar structure calculations, focusing initially on the resulting timescales for the magnetic activity cycles, and subsequently assessing the implications for the feasibility of the Applegate mechanism.

\subsection{Stellar structure and dynamo timescales}

As an example of our stellar structure calculations, Fig.~\ref{fig:dp-leo-conv-vel} shows the convection velocity and density profile for DP Leo as simulated with MESA. As we are considering rather rapidly rotating stellar systems as a result of tidal locking, we have explored the impact of varying the rotational velocity of the star, which is illustrated in Fig.~\ref{fig:dp-leo-conv-vel}, but find overall still moderate effects, with density changes of at most $20\%$ even in strong rotation scenarios.

\begin{figure}
\resizebox{\hsize}{!}{\includegraphics{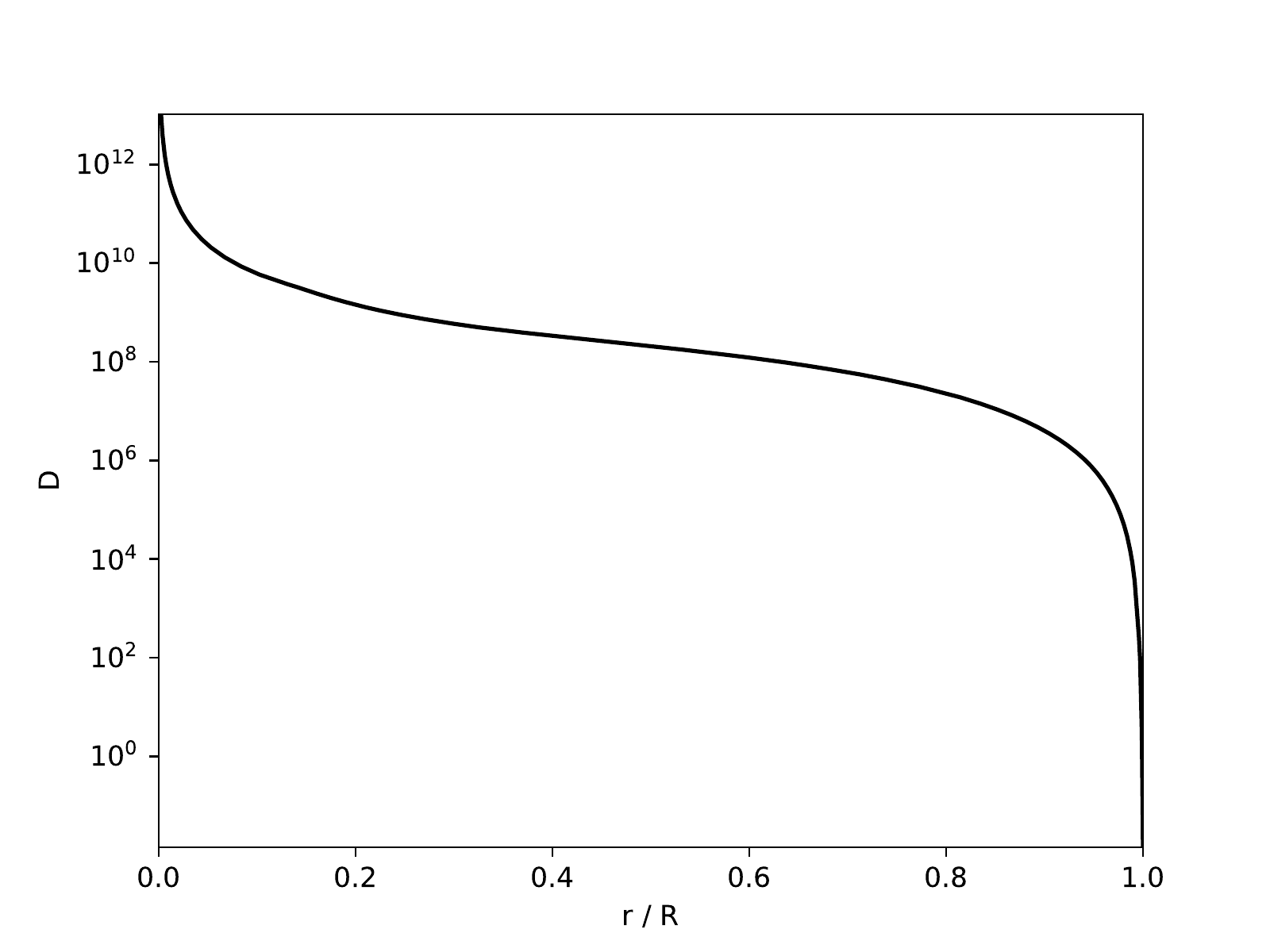}}
\resizebox{\hsize}{!}{\includegraphics{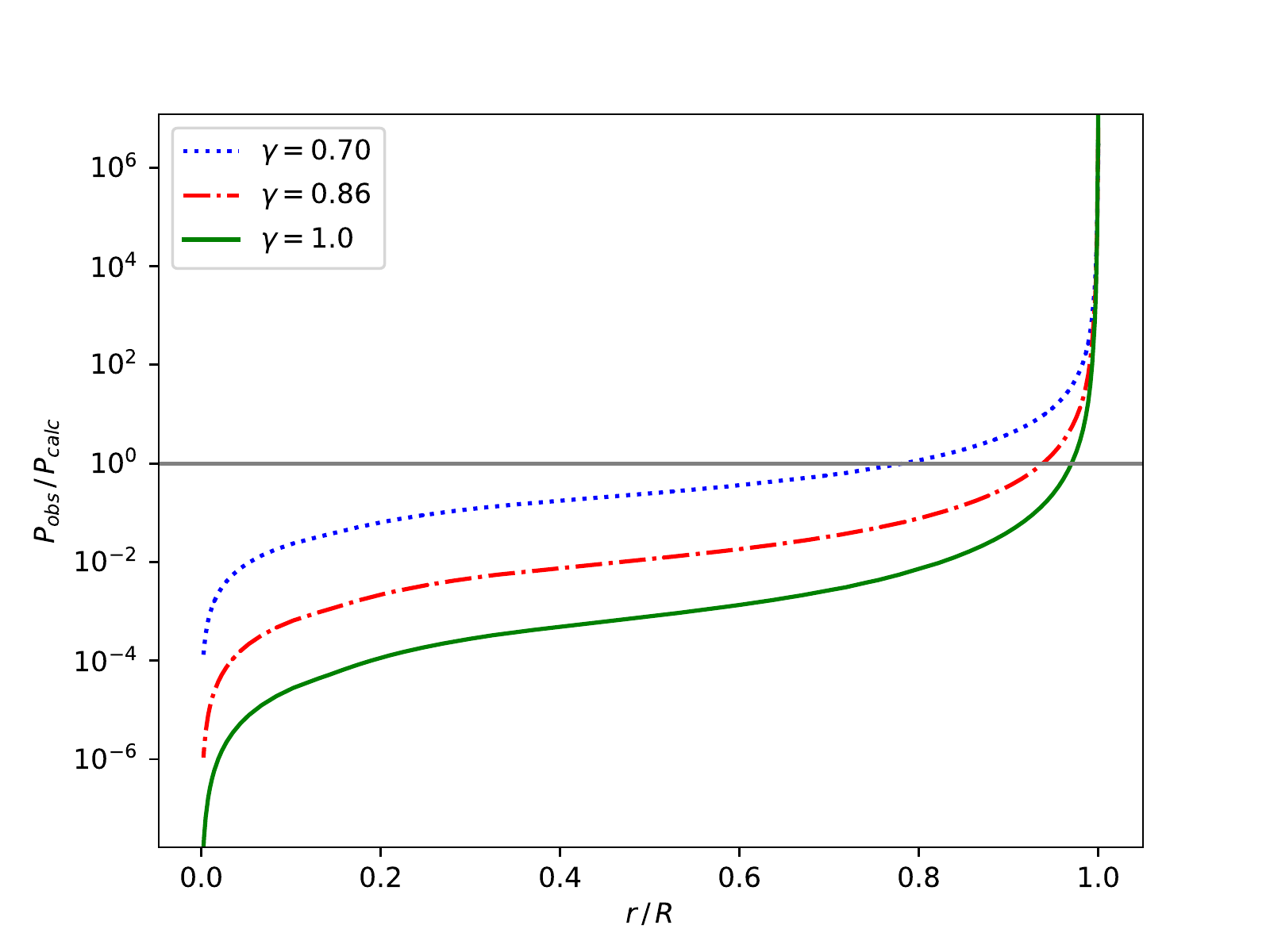}}
\caption{Dynamo number profile (top) and ratio of the observed vs. calculated activity cycle (bottom) for DP Leo calculated using Eq.~\ref{eq:period}.}
\label{fig:dp-leo-dynamo}
\end{figure}

Figure \ref{fig:dp-leo-dynamo} shows the dynamo number $D$ calculated as a function of radius for DP Leo based on eq.~(\ref{eq:ro}). With $\gamma = 0.86$ and $P_\textnormal{rot} = P_\textnormal{binary}$ we then calculated $P_\textnormal{cycle}$ using Eq.~(\ref{eq:period}) and the resulting dynamo number $D$. Figure~\ref{fig:dp-leo-dynamo} also shows the normalized period as a function of the radius. The horizontal gray line corresponds to a ratio of one, and the intersection thus marks the point where the calculated activity period equals the observed modulation period. This intersection point provides a first estimate for the effective radius that produces the observed modulation period. 

We have pursued the corresponding analysis for the $12$~PCEB systems  provided in the sample by \citet{voelschow16}, and summarize the obtained results together with the main parameters of the secondary in table~\ref{table:alpha}. In particular, we provide the calculated ratios $R_d/R$ for $\gamma=0.7$, $0.86$ and $1.0$. We generally find a reduced ratio of $R_d/R$ for a lower value of $\gamma$. Specifically, for $\gamma=0.7$, the ratios range between $0.6$ and $0.88$, for $\gamma=0.86$ they range between $0.78$ and $0.97$, and for $\gamma=1.0$ the ratios range between $0.87$ and $0.98$. While the correct value of $\gamma$ is not clear, we nevertheless note the general trend that the dynamo appears to be driven within the outer parts of the star, corresponding to radio of at least $60\%$ of the stellar radius. Only in case of V471~Tau, we have to note that the observed activity cycle cannot be reproduced for $\gamma=0.7$ or less, providing a stronger constraint on dynamo models for that system.

\begin{table*}
\caption{Main stellar parameters together with the relative rotation of the secondary star. Different radii $R_d$ are calculated for different $\gamma$ in Eq. \ref{eq:period} for which the rotation period $P_\textnormal{rot}$ is equal to the calculated activity period $P_\textnormal{cycle}$. The value $\gamma~=~0.86$ is that found for the Sun.}
\centering
\begin{tabular}{c c c c | c c c}
\hline
\hline
System & $M_\textnormal{sec}/M_\odot$ & $R_\textnormal{sec}/R_\odot$ & $\Omega/\Omega_\textnormal{crit}$ & & $R_d/R$ & \\
\hline
 & & & & $\gamma = 0.7$ & $\gamma = 0.86$ & $\gamma = 1.0$ \\
\hline
RX J2130.6+4710 & 0.555 & 0.534 & 0.116 & 0.67 & 0.86 & 0.92 \\
HS 0705+6700 & 0.134 & 0.186 & 0.265 & 0.87 & 0.95 & 0.97 \\
HW Vir & 0.142 & 0.175 & 0.192 & 0.60 & 0.89 & 0.95 \\
NN Ser & 0.111 & 0.149 & 0.154 & 0.81 & 0.96 & 0.97 \\
NSVS 14256825 & 0.109 & 0.162 & 0.208 & 0.88 & 0.94 & 0.98 \\
NY Vir & 0.15 & 0.14 & 0.155 & 0.88 & 0.96 & 0.98 \\
HU Aqr & 0.18 & 0.22 & 0.324 & 0.79 & 0.93 & 0.96 \\
QS Vir & 0.43 & 0.42 & 0.318 & 0.80 & 0.92 & 0.96 \\
RR Cae & 0.183 & 0.209 & 0.085 & 0.80 & 0.92 & 0.96 \\
UZ For & 0.14 & 0.177 & 0.262 & 0.82 & 0.94 & 0.97 \\
DP Leo & 0.1 & 0.134 & 0.288 & 0.78 & 0.93 & 0.97 \\
V471 Tau & 0.93 & 0.96 & 0.216 & - & 0.78 & 0.87 \\
\hline
\end{tabular}
\label{table:alpha}
\end{table*}


\subsection{Feasibility of the Applegate mechanism}\label{feasibility}

\begin{figure*}
\resizebox{\hsize}{!}{\includegraphics{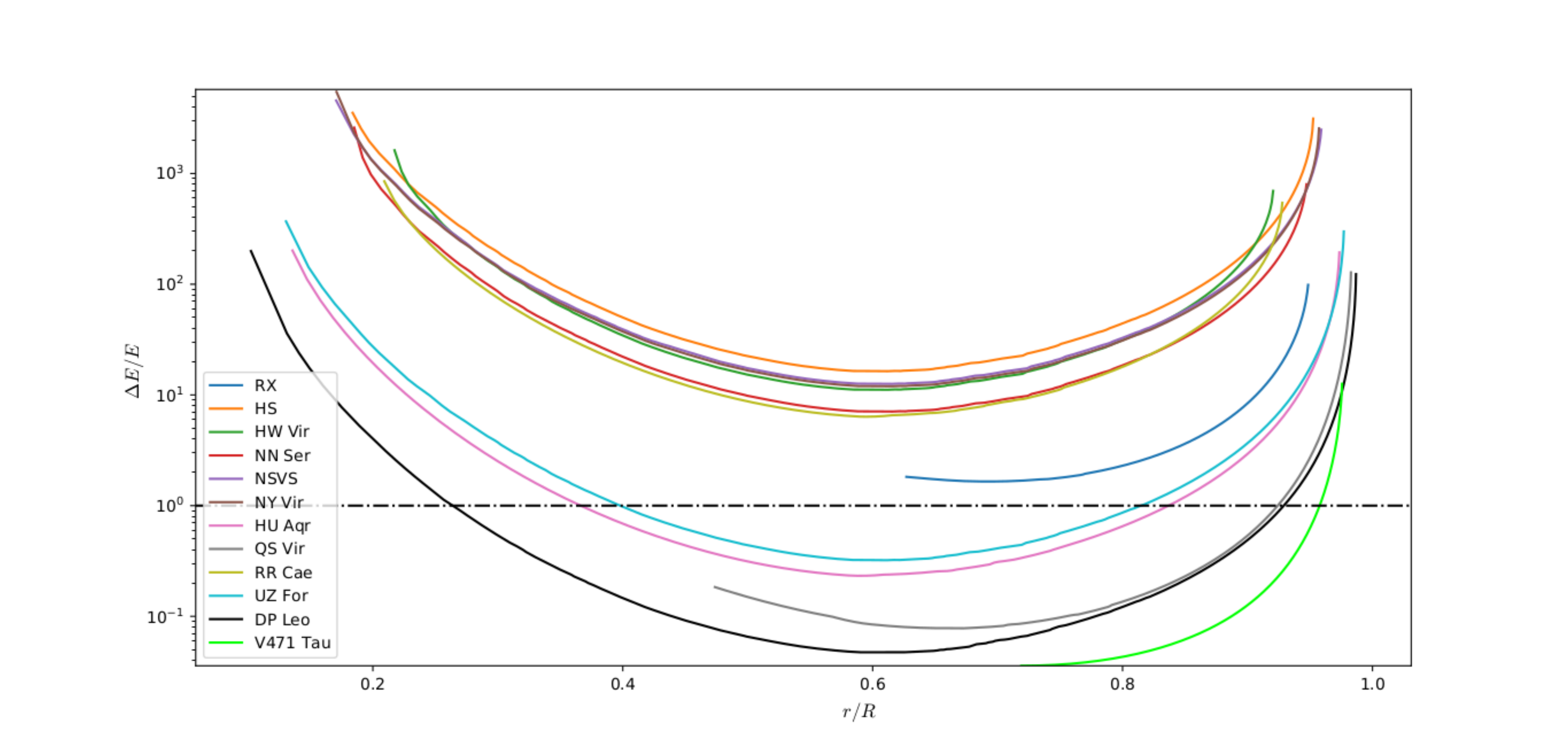}}
\caption{$\Delta E/E$ calculated for varying radii for all systems. Two groups can be identified, those that fell bellow $\Delta E/E = 1$ and those above it. At the borders the solutions for the Applegate energy become imaginary \citep[see][for a discussion on this]{voelschow16}.}
\label{fig:e_vs_r_all}
\end{figure*}

\begin{figure}
\resizebox{\hsize}{!}{\includegraphics{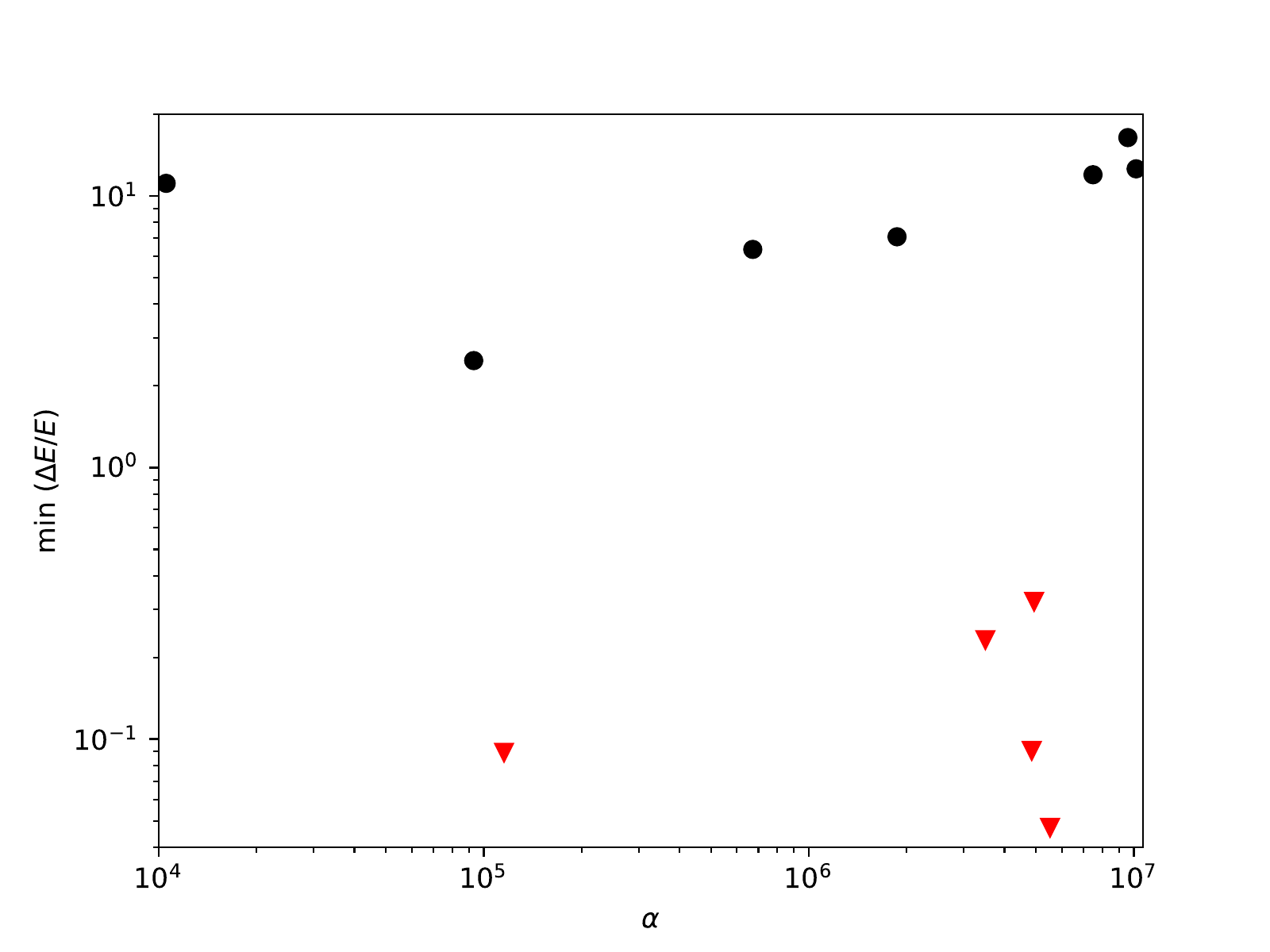}}
\resizebox{\hsize}{!}{\includegraphics{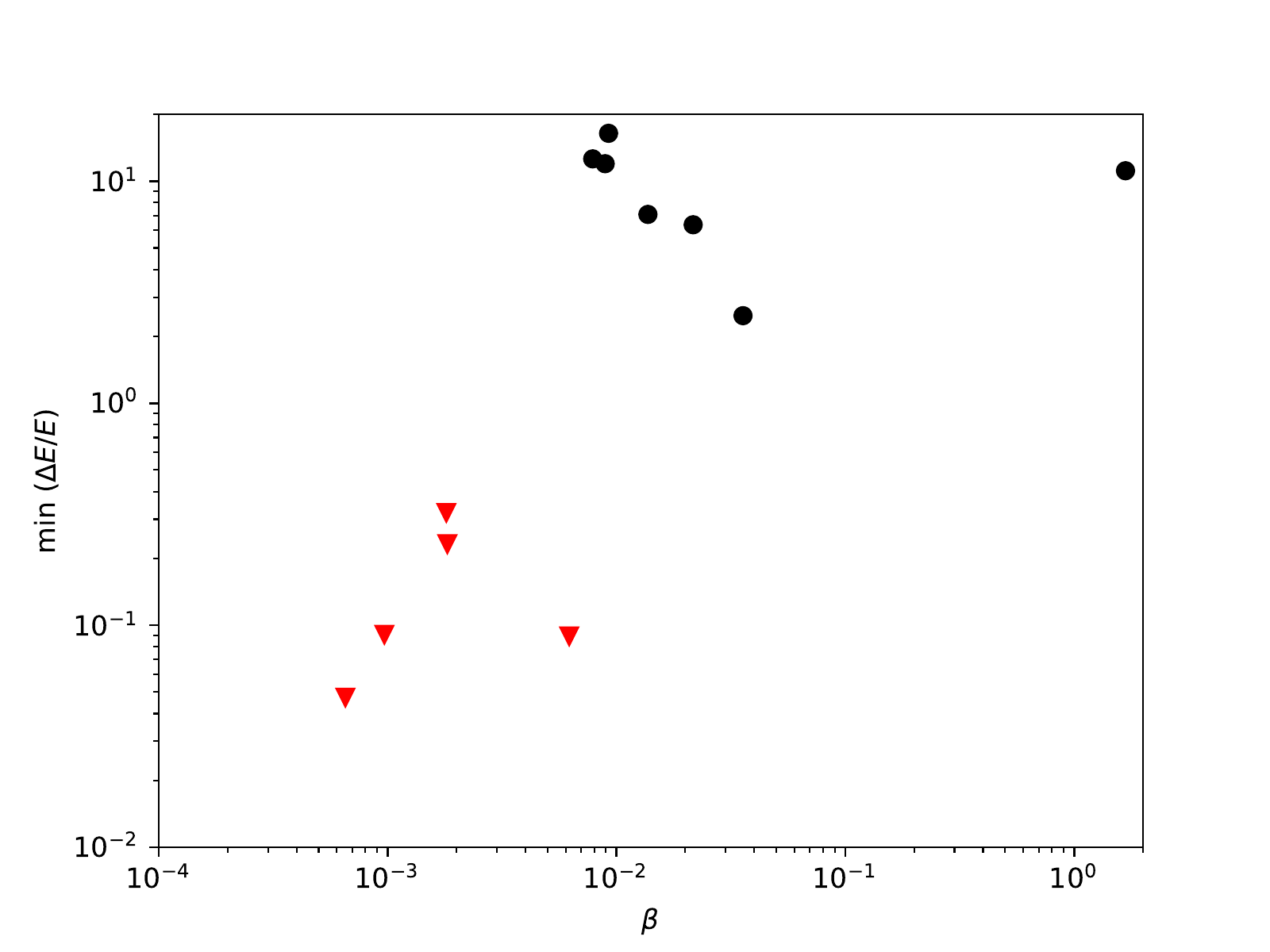}}
\caption{\textit{Top}: Minimum value of $\Delta E / E$ as a function of the parameter $\alpha$ defined in Eq.~\ref{alpha}. \textit{Bottom}: Minimum value of $\Delta E/E$ as a function of $\beta$ defined in Eq.~\ref{eq:beta}. Red triangles represent the systems falling below the horizontal dashed line at Fig. \ref{fig:e_vs_r_all}.}
\label{fig:alphabeta}
\end{figure}

\begin{figure}
\resizebox{\hsize}{!}{\includegraphics{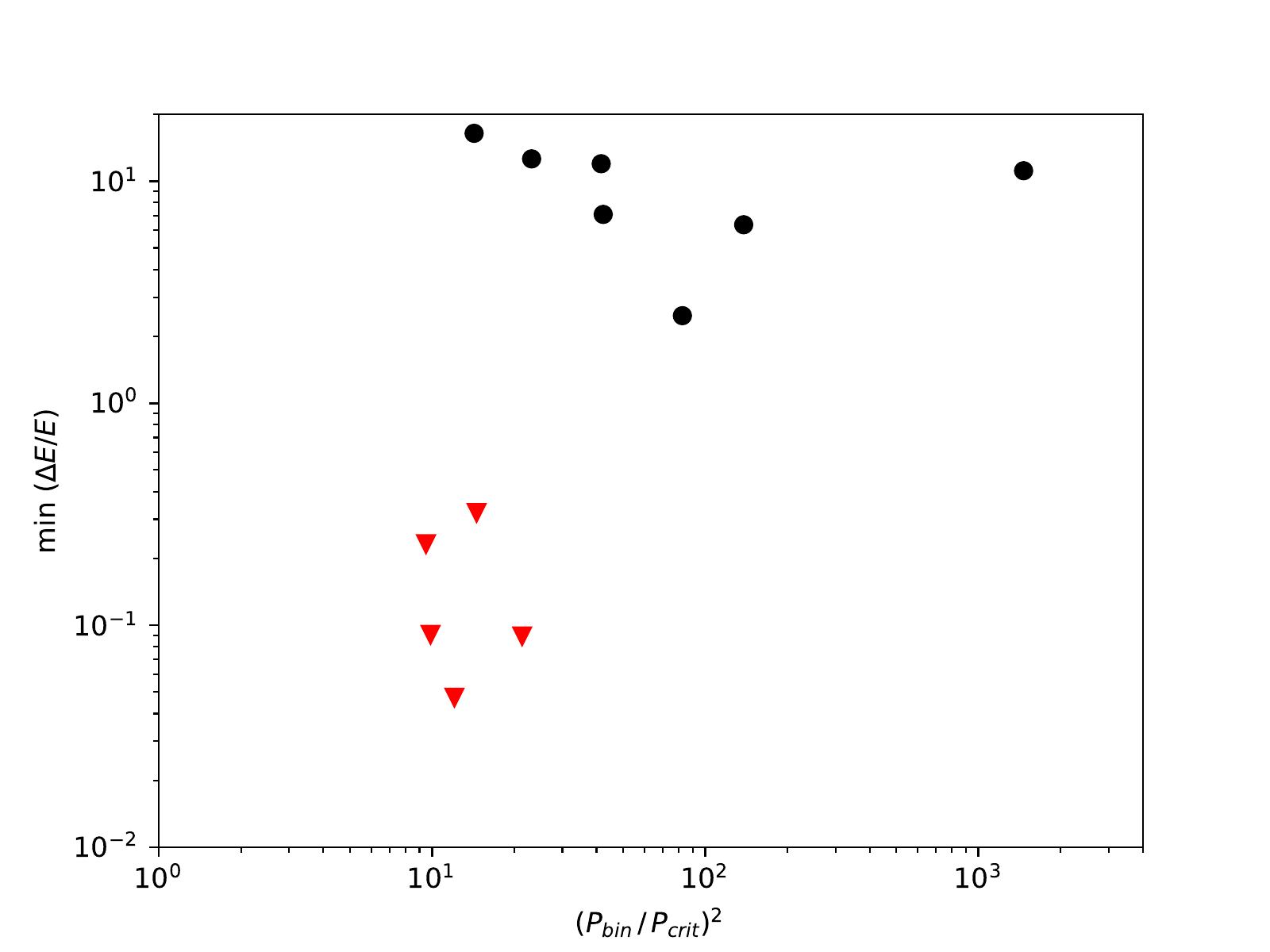}}
\resizebox{\hsize}{!}{\includegraphics{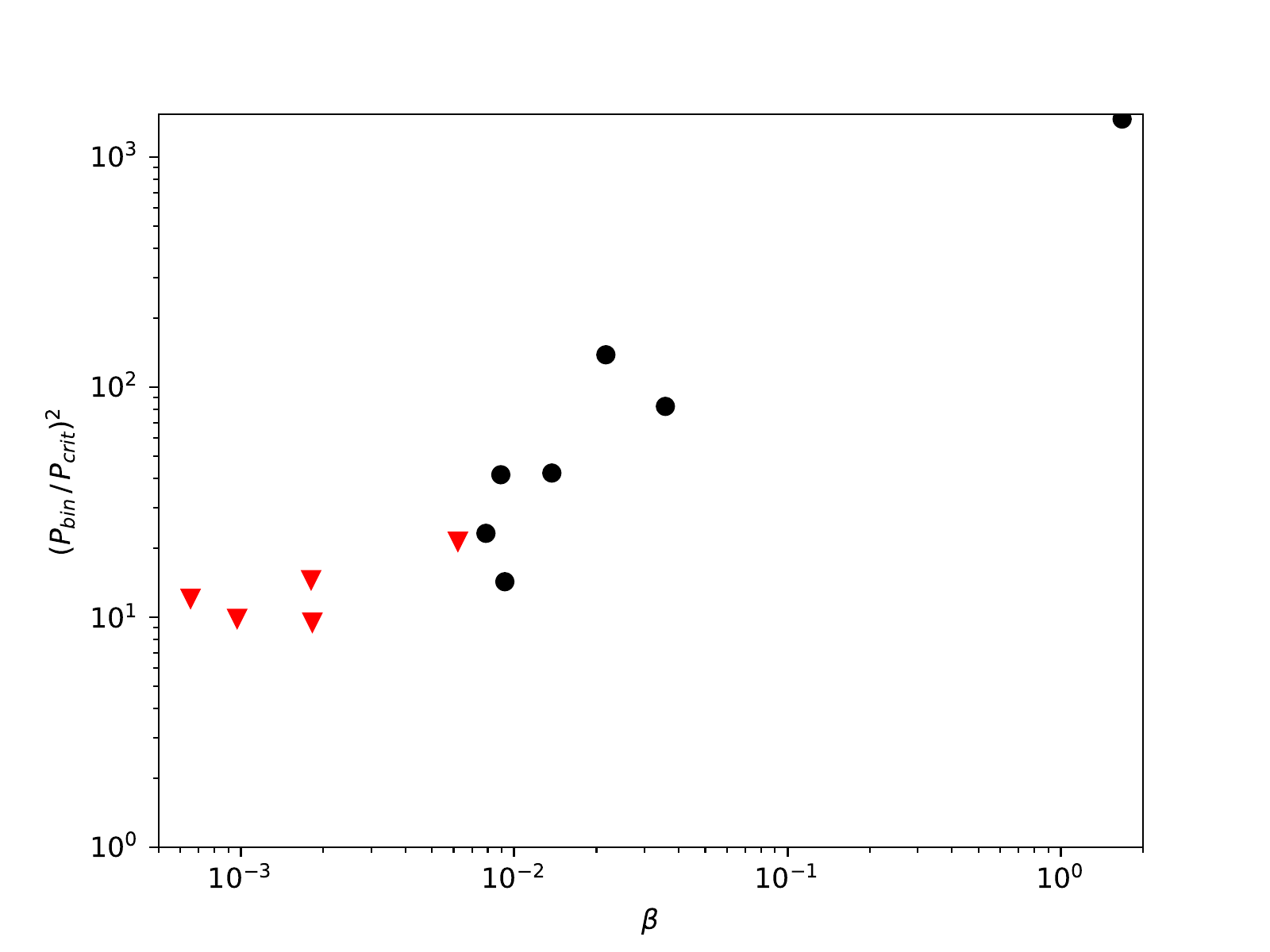}}
\caption{\textit{Top}: Minimum value of $\Delta E / E$ against the squared fraction between the binary period $P_\textnormal{bin}$ and the critical period $P_\textnormal{crit}$ defined as the period the binary would have if the secondary is rotating at its critical rotation velocity under the assumption of tidal lockment. \textit{Bottom}: $(P_\textnormal{bin}/P_\textnormal{crit})^2$ plotted against $\beta$ defined in Eq. \ref{eq:beta}; the rightmost circle corresponds to HW~Vir. Red triangles are the systems falling below the horizontal dashed line at Fig. \ref{fig:e_vs_r_all}.}
\label{fig:period-beta}
\end{figure}

\noindent We now assess whether the Applegate mechanism in these systems is capable of driving the observed eclipsing time variations. For this purpose, the ratio of required to available energy is calculated using Eq.~\ref{eq:energy}. The results of the calculation are summarized in Table \ref{table:1} for different scenarios. We note that our calculations from Table \ref{table:1} are different from those reported in \citet{voelschow16} two-zone model (their Table 4). While they determined the core radius based on radial density profiles obtained with Evolve ZAMS\footnote{Webpage Evolve ZAMS:\\http://www.astro.wisc.edu/~townsend/static.php?ref=ez-web} that were rescaled to the stellar radius, we have employed here the radial profiles obtained via MESA star. More importantly, we varied the position that is considered as the separation point between the core and the shell, and we also explore the effect of choosing this position according to the radius where the modulation period is reproduced by our dynamo model. The \citet{voelschow16} results are reobtained at $r/R\sim0.88$. To give an idea of the uncertainties in the energetics, we provide error bars on $\Delta E/E$ in case of RX~J2130.6+4710 in Table~\ref{table:1}, which are evaluated adopting the upper and lower limits for secondary mass and radius, respectively. The latter shows that the values of $\Delta E/E$ should be taken as indicative, but come with uncertainties of about a factor of two. A more accurate determination of stellar parameters would certainly help to reduce these errors, though we caution the reader that intrinsic uncertainties are also present within the two-zone approximation of the model, and further theoretical development will be necessary as well. Particularly in the case of RX~J2130.6+4710, we note that $\Delta E/E$ may still be less than one if the error are taken into account.

If we first focus on our reference case with $\gamma=0.86$ and adopt the radius $R_d$ as the radius that separates the shell from the core, we find that in three systems, QS~Vir, DP~Leo and V471~Tau, the Applegate mechanism is energetically feasible, in two of them (QS~Vir and DP~Leo) though  only marginally so. The system RX~J2130.6+4710 in principle yields values $\Delta E/E>1$. Considering the error bars, it may however be consistent still with the scenarios where Applegate is marginally feasible. We also note that, in case of RX~J2130.6+4710, there is only a lower limit available for the modulation period, translating into an upper limit on $\Delta E/E$. As we noted above, it is not clear that the normalization to $\gamma=0.86$ is appropriate for these type of systems, or if even a constant value of $\gamma$ is justified, given the different types of secondary stars in the sample. We thus explore how the energetic feasibility of our results depends on the adopted ratio of $R_{\rm core}/R$, exploring ratios of $0.85$, $0.75$ and $0.65$, which lie within the plausible range determined in the previous subsection. We find in particular that the required energy at least initially decreases when adopting a larger shell, only for some systems it starts increasing again at $R_{\rm core}/R=0.65$. Exploring the results in this parameter space, we then find that in five systems, HU~Aqr, QS~Vir, UZ~For, DP~Leo and V471~Tau, the Applegate mechanism is potentially feasible, with ratios $\Delta E/E$ ranging from $0.036$ to $0.98$. 

To illustrate this further, we plot $\Delta E/E$ as a function of $r/R$ in Fig. \ref{fig:e_vs_r_all}. The curves end when there is no longer a physically meaningful solution, that is, when the required change in the angular momentum in no longer consistent with angular momentum conservation within the star. For all systems, we see that the required energy fraction follows an approximately parabolic shape with a minimum near $R_{\rm core}/R\sim0.6$. While all curves are of similar shape and share the same asymptotic behavior, the magnitude or normalization varies significantly, and the systems can be roughly divided into two groups. In particular, there is a first group of six systems with the minimum $\Delta E/E\sim60-300$, where thus the Applegate mechanism is clearly not feasible, and another group consisting of the systems identified above, with minimum $\Delta E/E\sim 0.03-0.9$, where the mechanism becomes feasible. While the gap between these groups appears striking within the figure and corresponds to roughly two orders of magnitudes in $\Delta E/E$, we cannot entirely exclude in full that the later is due to the relatively small number of systems in our sample. While the system RX~J2130.6+4710 appers to lie somewhat in between the gap, though towards the lower end, it is important to note that the values of $\Delta E/E$ here represent only an upper limit, and better constraints on the modulation period will be necessary to evaluate whether the systems lies within the gap, or if it is part of the lower group. Now, within the error bars, the system is now already consistent with potentially driving the Applegate mechanism.

In order to determine if there is a physical origin of the bimodality that is indicated in the plot, we examined whether the minimum value of $\Delta E/E$ depends critically on relevant physical parameters of the system. To do so, we introduced the parameters
\begin{equation}
\alpha=k_1\,\frac{M_\textnormal{sec}R_\textnormal{sec}^2}{P_\textnormal{bin}^2P_\textnormal{mod}L_{sec}}\label{alpha}
\end{equation}
and
\begin{equation}\label{eq:beta}
 \beta= k_2G\frac{a_\textnormal{bin}^2M_\textnormal{sec}P_\textnormal{bin}^2}{R_\textnormal{sec}^5}\frac{\Delta P}{P_\textnormal{bin}}
,\end{equation}
from our Eq.~\ref{eq:energy}. In Fig.~\ref{fig:alphabeta}, we plot $\Delta E/E$ as a function of both parameters. While $\Delta E/E$ shows no correlation with $\alpha$, it is clearly recognizable that the lowest values of $\Delta E/E$ can be found for the lowest values of $\beta$ with a clear trend, with the remaining scatter being introduced by the dependence on $\alpha$. The latter provides a first indication that the parameter $\beta$ is most relevant in determining the feasibility of the Applegate mechanism.

While the parameter $\alpha$ shows no correlation with $\Delta E/E$, we show in  Fig. \ref{fig:period-beta} that the minimum value of $\Delta E / E$ clearly depends on the parameter $\beta$, with the remaining scatter being due to the dependence on $\alpha$. In the following, we also consider the ratio of binary period vs critical period of the secondary star, where the critical period is defined through the breakup velocity when the star reaches Keplerian rotation, i.e.,
\begin{equation}
 P_\textnormal{crit}^2 = 4\pi^2\frac{R_\textnormal{star}^3}{GM_\textnormal{star}},
\end{equation}
thus
\begin{equation}\label{eq:p_div_pcrit}
 \left(\frac{P_\textnormal{binary}}{P_\textnormal{critic}}\right)^2 = \frac{M_\textnormal{sec}}{M_\textnormal{total}}\left(\frac{a_{\rm bin}}{R_\textnormal{sec}}\right)^3 .
\end{equation}
Under the assumption of tidal locking, which is highly plausible within these systems, this ratio thus describes the ratio of actual over critical rotation period of the secondary star. We show in Fig.~\ref{fig:period-beta} how minimum $\Delta E/E$ depends on this quantity. In this case, the correlation is less clear as the correlation with $\beta$, but it is nevertheless visible that low ratios of $\Delta E/E$ occur only for low ratios of $P_{\rm binary}/P_{\rm critic}$, thus implying that high rotation rates (relative to breakup) are beneficial to drive the Applegate mechanism. In the same figure, we also explore whether the parameter $\beta$ is correlated with $P_{\rm binary}/P_{\rm critic}$, and also here we find that low values of $P_{\rm binary}/P_{\rm critic}$ imply low values of $\beta$, and vice versa. The latter strengthens our conclusion that rapid rotation may play a central role in the feasibility of the Applegate mechanism in PCEB systems.

In fact, one can show that \begin{equation}\label{eq:beta2}
\beta=k_2\frac{4\pi^2 a_{\rm bin}^2}{R_{\rm sec}^2}\frac{P_{\rm bin}^2}{P_{\rm critic}^2}\frac{\Delta P}{P_{\rm bin}},
\end{equation}
suggesting a quadratic dependence on the ratio of rotation period over critical period. We further note here that rapid rotation may frequently imply a short binary period and thus also a lower value of $a_{\rm bin}$, which strengthens this correlation further. The remaining parameter $R_{\rm sec}$ depends primarily on the mass of the secondary star, which is very similar in many of the PCEB systems (see Table~\ref{table:alpha}), and only occasionally introduces scatter when the secondary star is more massive. Equation \ref{eq:beta2} can be rewritten using Eq. \ref{eq:p_div_pcrit}, leading to
\begin{equation}
 \beta = k_2 4\pi^2 \left(\frac{P_\textnormal{bin}}{P_\textnormal{critic}}\right)^{10/3}\left(\frac{M_\textnormal{total}}{M_\textnormal{sec}}\right)^{2/3}\frac{\Delta P}{P_\textnormal{bin}}.
\end{equation}
This last equation provides evidence of a stronger dependence on the ratio of the rotation period over critical period. It is then plausible overall that the main visible dependence is due to rotation.

\begin{table*}
\caption{Results for the energy required to drive the Applegate mechanism for all systems considered in this work. $R_d/R$ denotes the core-shell transition radius. $\Delta E/E_\textnormal{sec}$ is the necessary energy to drive the change in the quadrupole moment to give raise to the observed period, as a fraction of the available energy, calculated considering different shells including the point at which the observed period matches the calculated, i.e., $R_d/R$. Results with $\Delta E / E_\textnormal{sec} < 1$ are highlighted in bold. RX~J2130.6+4710 is presented with errors, as mentioned in section \ref{feasibility}}.
\centering
\begin{tabular}{c | c c c c c}
\hline\hline
System & & & $\Delta E/E_\textnormal{sec}$ & &  \\    
\hline
 & $R_d/R = R_d/R\,(\gamma = 0.86) $ & $R_d/R = 0.85$ & & $R_d/R = 0.75$ & $R_d/R = 0.65$ \\
\hline
RX J2130.6+4710 & 4.2$^{-2.3}_{+7.6}$ & 3.6$^{-1.8}_{+5.4}$ & & 1.8$^{-0.96}_{+2.5}$ & 1.7$^{-0.93}_{+2.3}$ \\
HS 0705+6700 & 1354 & 56 & & 22 & 93 \\
HW Vir & 93 & 43 & & 15 & 16  \\
NN Ser & - & 24 & & 9.5 & 6.7 \\
NSVS 14256825 & 500.5 & 57 & & 21 & 13 \\
NY Vir & - & 54 & & 19 & 12 \\
HU Aqr & 6.3 & \textbf{0.88} & & \textbf{0.41} & \textbf{0.25} \\
QS Vir & \textbf{0.88} & \textbf{0.22} & & \textbf{0.097} & \textbf{0.078} \\
RR Cae & 237.3 & 34 & & 11 & 6.8 \\
UZ For & 12 & 1.5 & & \textbf{0.56} & \textbf{0.34} \\
DP Leo & \textbf{0.98} & \textbf{0.21} & & \textbf{0.081} & \textbf{0.049} \\
V471 Tau & \textbf{0.039} & \textbf{0.061} & & \textbf{0.036} & - \\
\hline
\end{tabular}
\label{table:1}
\end{table*}


\section{Discussion of selected systems}\label{comparison}

\subsection{HW Vir}
Consisting of a primary of $0.485$~M$_\odot$ and a secondary of $0.142$~M$_\odot$, this system was first proposed to have a two-planet system by \citet{lee09} that was later proven to be secularly unstable and replaced with another two-planet secularly-stable system by \citet{beuermann13b}. We find rather high values of $\Delta E / E$ for this system (see Figure \ref{fig:e_vs_r_all}), and thus HW~Vir might be a good candidate to test the planetary hypothesis using direct imaging.

\subsection{NN Ser}
This system consists of a white dwarf with a M4 companion \citep{parsons10}. The planetary solutions that explain the eclipsing time variations were proven to be dynamically stable \citep{beuermann13a}. This, together with the recent detection of dust around NN Ser by \citet{hardy16} that was not expelled from the common envelope phase, add credibility to the planetary hypothesis. On the other hand, the most recent data require the subtraction of a quadratic term from the ephemeris to obtain the planet solution \citep{bours16}. Based on the calculations in this paper, we find that it is energetically difficult to explain the observed eclipsing time variations based on the Applegate effect alone, as there are no solutions with $\Delta E/E<1$. However, there are solutions with $\Delta E/E\lesssim10$, suggesting that relevant fluctuations of the eclipsing time variations could be induced by the Applegate mechanism, even if it is not the entire signal. 

\subsection{HU Aqr}
HU Aquarii is a system of particular interest. First discovered by \citet{schwope93}, it has a strongly magnetized primary white dwarf and a M4 secondary and it is constantly beeing monitored by several groups. \citet{bours14} concluded that a planetary model with up to three members cannot explain the observed ETVs as they are dynamically unstable over short periods of time. \citet{gozdiewski15} revisited the planetary hypothesis and concluded that it might be possible that a three-planet system is present. However, either one of these planets must have a retrograde orbit or they should have high mutual inclinations.

In our calculations, HU Aqr was found to indeed have solutions that might trigger the Applegate mechanism, with $\Delta E/E\sim0.3$. Such solutions require the dynamo to operate further in the interior of the star. The ratio of $\Delta E/E$ is reduced compared to the value reported by \citet{voelschow16} due to the the evaluation of the coefficients $k_1$ and $k_2$ from the stellar structure profiles.

\subsection{QS Vir}
\citet{horner13} performed a detailed dynamical study of the proposed planets for QS Vir. The authors performed more than 180~thousand simulations of the proposed planetary systems but found that none of them with long enough stability to explain the eclipsing time variations. More recently, \citet{parsons16} used high-resolution spectroscopy to study the magnetic activity of this PCEB and found that the M dwarf is covered with a large number of star spots, thus indicating that it is a very active star.

\indent We have found that the M dwarf on this system is among those with the highest relative rotation, where the Applegate mechanism appears to be feasible (see Table \ref{table:alpha}). The secondary dM star has a radiative core, which we assumed here not to contribute to the Applegate mechanism.

\subsection{V471 Tau}
It was predicted by \citet{beavers86} that a third body with a period of $\sim25$~years could exist in this system. Recently, \citet{Hardy15} used direct imaging to test the hypothesis, resulting in a nondetection of the proposed brown dwarf. Later, \citet{vaccaro15} analyzed the system and proposed several ways for the third body to have avoided detection. However, \citet{vanderbosch17} studied the eclipsing time variations with two independent clocks, namely the orbital period of the binary and white dwarf spin period. They do not find the same magnitude of variation of the O-C of the spin period as for the diagram based on the binary period. The latter thus favors the Applegate mechanism as opposed to the presence of a third body.

As the secondary of V471~Tau is a Sun-like star with a mass of $\sim 0.93\,$M$_\odot$, it is expected to have an $\alpha\Omega$ dynamo operating on the radiative-convective interphase. As this star is rotating much faster than the Sun, one may expect that a strong magnetic field due to rapid rotation may trigger the Applegate mechanism, consistent with our findings.

\subsection{NSVS 14256825}
For the system NSVS 14256825, the cyclic behavior of the O-C residuals was previously attributed to the presence of one or two Jupiter-like planets \citep{Beuermann12, Almeida13, Wittenmyer13}, new data by \citet{nasiroglu17} revealed a systematic, quasi-sinusoidal variation deviating from the older ephermis by about $100$~s. As the most plausible explanation for this deviation, they propose a one-companion model to be the most reliable explanation and propose that the Applegate mechanism is not energetically feasible to drive the changes. We note here that while there are no solutions with $\Delta E/E<1$ that could explain the eclipsing time variations entirely, there are solutions with $\Delta E/E\sim10$, implying that magnetic activity could at least induce relevant scatter in the observed variations.

\subsection{RX J2130.6+4710}
This PCEB consists of a red dwarf with a mass of $0.555~M_\odot$ and radius $0.534~R_\odot$, and a white dwarf of mass $0.554~M_\odot$ and radius $0.0137~R_\odot$ with a separation of 2-3 $R_\odot$ \citep{maxted04}. The large O-C variations from a relatively long observational baseline were detected by \citet{bours16}. They identified this system to have a lower limit for the period of the mechanism at work to be 30 yr and we adopted this limit for our calculation together with an eclipsing time variation of $\sim 250$ s. This translates into $\Delta P/P = 1.6\times10^{-6}$. Considering the error bars, the system may be able to drive the Applegate mechanism. In addition, if the modulation period is greater, say 50~yr, then $\Delta P/P \sim 9\times10^{-7}$ and the system will clearly fall bellow the line $\Delta E/E = 1$. As noted by \citet{bours16}, the observational baseline is yet too small to draw a more robust conclusion for this system and further eclipsing observations will be crucial.


\section{Conclusions}\label{conclusions}

In this paper we have explored whether the Applegate mechanism can drive the observed eclipsing time variations in a sample of $12$ PCEB systems. For this purpose, we have obtained radial profiles of the secondaries using the MESA code \citep{mesa11}, and applied simple dynamo models following \citet{Soon93} and \citet{Baliunas96} to determine the radius at which the expected activity cycle matches the observed modulation period. This radius was found to depend on the power-exponent $\gamma$, but even for different choices of $\gamma$, we found that the dynamo can generally be expected to be driven in the outer layers, at radii $R_d/R\gtrsim0.6$.

We subsequently explored whether the Applegate mechanism to drive the eclipsing time variations is also energetically feasible, employing the framework presented by \citet{voelschow16}, which is now publicly available through the  \texttt{Applegate calculator}\footnote{Applegate calculator:\newline http://theory-starformation-group.cl/applegate/}. Based on this analysis, we show that the Applegate mechanism may be energetically feasible in up to five systems,  HU~Aqr, QS~Vir, UZ~For, DP~Leo and V471~Tau. Plotting the required energy $\Delta E/E$ as a function of the assumed radius of the core, we also find that the latter yields a characteristic shape for all systems explored here, with a minimum at a core radius $R_{\rm core}/R\sim0.6$. The normalization of the curves depends however on the system, which appears to be separated into two groups, those with $\Delta E/E\sim0.03-0.9$ where the Applegate mechanism is clearly feasible, and another group with $\Delta E/E\sim60-300\gg1$. Our current sample shows a gap in between these groups of about two orders of magnitude in $\Delta E/E$. It does, however, remain to be explored whether the gap is of physical origin. In the case of RX~J2130.6+4710, whether or not it is part of the lower group, with $\Delta E/E < 1$, or if it indeed lies within the gap needs to be further explored. Such a determination is currently not possible, as there is only a lower limit on its modulation period, and observations over a longer baseline will be necessary to test this possibility.

We explored further what determines if the Applegate mechanism is feasible in a particular system, focusing on the parameters $\alpha$ and $\beta$ in our Applegate framework. While the first parameter appears uncorrelated to the minimum $\Delta E/E$, $\beta$ shows a clear correlation for our sample. We also show that this correlation appears related to the strength of rotation in the system, particularly the ratio of the rotation period of the secondary star (which is equal to the binary period in the case of tidal locking) over the critical breakup period of the secondary. In particular, we emphasize that all systems with low minimum values of $\Delta E/E$ have low ratios of $P_{\rm bin}/P_{\rm critic}$. The latter suggests that rotation plays a relevant role in determining the feasibility of the Applegate mechanism. The effect that rapid rotation plays on the magnetic field of M dwarfs is explored by \citet{morgan12} using WD+dM binaries and comparing the activity level and strength to field dM. The authors find that dM have a higher activity fraction in the range of spectral types M$0$ to M$7$, when they are in a close binary and it was attributed to the increment of stellar rotation in the paired dMs. In the same study, it was also found that the activity is increased in WD+dM across all spectral types when compared to the unpaired dM. This strengthens our main conclusion, namely higher rotation relative to critical rotation supports the Applegate mechanism PCEBs.

We also note that there may be other mechanisms that can change the quadrupole moment of a star. A minor effect could be related to the side of the M dwarf facing the white dwarf, which will be more deformed than the other side, and this effect is expected to depend on the binary separation. The magnetic fields from the white dwarfs in the systems considered here have been neglected so far, but it is also straightforward to show that their expected contributions would be smaller than typical surface fields from M dwarfs required to drive the Applegate mechanism \citep[see][]{applegate92}. A generic remark that is important to make is that several of the systems examined here cannot be explained under the Applegate hypothesis (at least with the two-zone model adopted here), and it remains to be explored if another mechanism is operating in them, or if the latter hints at a more fundamental problem in our current understanding.

To make further progress on this topic, we encourage direct imaging attempts as pursued by \citet{Hardy15}, particularly for cases where the presence of planets appears plausible. In addition, other methods like radial velocity measurements can be employed to determine independent constraints on the potential motion of the secondary star \citep{Osagh17}. The precise photometry obtained via the Gaia satellite\footnote{Webpage Gaia: http://sci.esa.int/gaia/} may further be valuable to determine if motions of the center of mass of the systems can be confirmed in the plane of the sky. Such independent constraints or measurements will be extremely valuable to confirm or rule out the existence of planets. 

At the same time, it will be important to probe the presence of magnetic activity and in particular the duration of the magnetic cycle, as recently pursued for instance by \citet{Perdelwitz17}. The latter is critical to understanding if the observed modulation period is consistent with the activity cycle of the secondary star, and provides an independently relevant test to probe the physics of the eclipsing time variations. We also note here that stellar magnetism and the presence of planets are not necessarily mutually exclusive alternatives, but may simultaneously occur. In particular, we have identified several systems here where we expected that magnetic activity may not explain the observed eclipsing time variations, but include relevant scatter in the observed eclipsing times, due to values of $\Delta E/E\sim10$, corresponding to fluctuations on the $10\%$ level. The presence of such fluctuations may potentially also explain that in some cases, the subtraction of approximately quadratic terms is needed, as noted recently by \citet{bours16} in the case of NN~Ser. They also presented eclipsing times for a set of 67 close binaries. Besides the PCEBs from \citet{voelschow16} we have included RX J2130.6+4710. This is because this is the only new system with a long enough observational baseline to constrain the modulation period, namely at least 30 yr, and with large O-C variations. Extending the baselines will be important to extend the analysis presented in this paper.

We expect additional theoretical work to be necessary in the future, as the Applegate model employed here \citep{voelschow16}, while being a significant improvement compared to the first version laid out by \citet{applegate92}, is still based on a two-zone approximation. While some suggestions for generalizations exist \citep[e.g.,][]{Lanza05, Lanza06}, the models require further exploration before applying them systematically to a larger sample of systems, and need to be linked more strongly to the dynamo itself. We expect this to be a promising area for further developments.

\begin{acknowledgements}
     The authors thank the referee for useful comments that helped to improve this work. We are grateful for funding through Fondecyt regular (project code 1161247). DRGS is grateful for funding through the ''Concurso Proyectos Internacionales de Investigaci\'on, Convocatoria 2015'' (project code PII20150171) and ALMA-Conicyt (project code 31160001). FHN, JZ and DRGS further acknowledge funding from BASAL Centro de Astrof\'isica y Tecnolog\'ias Afines (CATA) PFB-06/2007.
\end{acknowledgements}

\bibliographystyle{aa} 
\bibliography{pceb.bib} 

\end{document}